\documentclass[a4paper,fleqn]{cas-dc}
\usepackage[numbers]{natbib}
\usepackage{amsmath, amssymb, amsthm}
\usepackage{placeins}
\usepackage{graphicx}
\usepackage{booktabs}
\usepackage{hyperref}
\usepackage{longtable}
\usepackage{booktabs} 
\usepackage{multirow}
\usepackage{pifont}
\usepackage[font=small,labelfont=bf]{caption} 
\usepackage{tabularx}
\usepackage{float}
\usepackage[table]{xcolor}
\usepackage{adjustbox}

\def\tsc#1{\csdef{#1}{\textsc{\lowercase{#1}}\xspace}}
\tsc{WGM}
\tsc{QE}

\begin{document}
\let\WriteBookmarks\relax
\def\floatpagepagefraction{1}
\def\textpagefraction{.001}

\shorttitle{Asymmetric Multimodal Fusion}    

% Short author
\shortauthors{Kou et al.}

% Author 1 - Equal contribution
\author[1]{Zijing Xu}[]
\fnmark[1]
\ead{2025223045139@stu.scu.edu.cn}
\credit{}

% Author 2 - Equal contribution  
\author[2]{Yunfeng Kou}[]
\fnmark[1]
\ead{2025226040003@stu.scu.edu.cn}
\credit{}

% Author 3 - Corresponding author
\author[1]{Hong Liu}[]
\cormark[1]
\ead{liuhong@scu.edu.cn} 
\credit{}

% Author 4
\author[1]{Kunming Wu}[]
\ead{2025223040059@stu.scu.edu.cn}  
\credit{}

\fntext[1]{These authors contributed equally to this work.}
\cortext[1]{Corresponding author}

% Main title of the paper
\title [mode = title]{Contribution-Guided Asymmetric Learning for Robust Multimodal Fusion under Imbalance and Noise}  

% Title footnote mark
% eg: \tnotemark[1]
% \tnotemark[1] 
% Title footnote 1.
% eg: \tnotetext[1]{Title footnote text}
% \tnotetext[1]{} 

% Address/affiliation
\affiliation[1]{organization={Department of Computer Science, Sichuan University},
            addressline={No. 24 South Section 1, First Ring Road}, 
            city={Chengdu},
            postcode={610065}, 
            state={Sichuan},
            country={China}}
\affiliation[2]{organization={National Key Laboratory of Fundamental Science on Synthetic Vision, Department of Computer Science, Sichuan University},
            addressline={No. 24 South Section 1, First Ring Road}, 
            city={Chengdu}, 
            postcode={610065},
            state={Sichuan},
            country={China}}

% Here goes the abstract
\begin{abstract}
Multi-modal learning faces two major challenges: modality imbalance and data noise, which significantly impact the robustness and generalization ability of models. Existing approaches, which enforce modality balancing by suppressing high-performing modalities, may lead to convergence to suboptimal solutions. In this paper, we propose a comprehensive contribution-based calculation method and a modality compression paradigm called CAL. This mechanism evaluates both marginal performance and dynamic potential, reinforcing the gradients of high-contribution modalities to achieve an adaptive balance between modalities. Asymmetric compression reduces redundancy in modality-specific information, mitigates noise, and strengthens the contribution of weak modalities, thus continuously optimizing multi-modal information fusion performance.

On five benchmark datasets, including sentiment analysis, scene recognition, and event localization tasks, CAL demonstrated outstanding performance in both imbalanced fusion tasks and noise robustness tests. In experiments on CREMA-D, KS, and AVE, the model achieved accuracy rates of 79.30\%, 74.82\% and 74.21\% respectively. Surpassing the current state-of-the-art model ARL. In high-noise robustness tests, our work also outperforms existing methods under various attack strategies on the MVSA-Single and NYUD2 datasets. These results validate the advantages of CAL in addressing modality imbalance and noise interference. As a flexible and efficient framework, it is easily transferable to other tasks, offering broad adaptability and potential for future applications.
\end{abstract}

% Use if graphical abstract is present
%\begin{graphicalabstract}
%\includegraphics{}
%\end{graphicalabstract}

% Research highlights
\begin{highlights}
\item Proposed a unified paradigm for addressing modality imbalance and data noise through intelligent information compression.
\item Proved the effectiveness of the modality enhancement approach and introduced a valid modality contribution calculation method.
\end{highlights}

%\nocite{*}

% Keywords
\begin{keywords}
Multimodal Learning \sep Modality Imbalance \sep Noise Robustness \sep Information Bottleneck \sep Gradient Modulation
\end{keywords}

\maketitle

% Main text
\newpage
\section{Introduction}

Multi-modal learning has emerged as a significant research direction in the field of artificial intelligence in recent years, aiming to improve model performance by integrating data from different modalities. In the research of multi-modal machine learning, ideal models often assume that modality information is balanced, reliable, and all input data is of high precision. However, recent work has shown that multi-modal models are often plagued by the issue of modality imbalance, with significant performance differences between modalities and incomplete reliability of the information \cite{wang2020makes, huang2022modality}. Additionally, multi-modal data is frequently subject to noise and interference \cite{yang2024quantifying, zeng2022robust}, and attacks on the dominant modalities, such as modality blurring, result in a significant drop in performance. These challenges related to modality utility and data quality severely limit the robustness and reliability of multi-modal systems in real-world, complex environments, and represent key issues that need to be addressed in current research.

Many works have proposed constructive solutions from perspectives such as model architecture \cite{zhao2021missing, li2023graphmft, zhang2024multimodal} and gradient adjustment \cite{wang2020makes, fan2023pmr}. However, existing methods still have certain limitations. On one hand, many robust fusion methods require additional callable modules \cite{gao2024embracing} or complex optimization objectives, which increase model complexity and computational overhead, making theoretical analysis and the transfer of models to other tasks difficult. On the other hand, in existing gradient adjustment methods, the mainstream idea \cite{peng2022balanced, fan2023pmr} is to suppress the optimization process of the dominant modality in order to create more training space for the weaker modalities. While modality suppression is an objectively existing phenomenon, the identification of the dominant modality remains problematic. ARL \cite{Wei2025ARL} has shown that the identification of dominant modalities should not be based solely on discriminating model performance; Reinforcing high-performing modalities might yield better results. It is important to acknowledge the inherent differences between modalities due to their information redundancy and intrinsic task relevance. The oversight of these differences in past methods may prevent models from fully exploiting the informational advantages of dominant modalities and even lead to convergence at a local optimum.

This paper, while ensuring the simplicity and ease of transferability of the method, designs a learning paradigm that can adapt to changes in data quality and intelligently balance the inherent value of different modalities. The paper does not deny the dominance of high model performance, but argues that it should not be the sole criterion for evaluation; The training potential of modalities should also be considered. This work intelligently compresses modality information while adjusting gradients, aiming to address both modality imbalance and data interference simultaneously. Specifically, this paper proposes a new multi-modal learning framework, which consists of two collaborative working mechanisms.

Strong modalities with high initial performance should not be suppressed but dynamically balanced in their reinforcement levels. The proposed method evaluates the real-time contribution of each modality based on both marginal performance and dynamic potential, and adaptively adjusts the gradient update magnitude for each modality, ensuring that all modalities are updated at relative speeds.

To tackle noise interference in multi-modal data and encourage the training of weak modalities, this paper introduces the Information Bottleneck theory as a guiding principle. The original features of each modality are mapped to a latent representation space, and the fused features are used as the compression target to ensure the validity of the compression direction. This mechanism compresses the features of all modalities, but the compression ratio is not fixed. A higher compression ratio is applied to modalities with low contribution or noise to suppress redundant information and enhance their contribution. This differentiated compression strategy aims to force the model to focus on the most discriminative parts of each modality, thus achieving more efficient and robust multi-modal fusion.

The method proposed in this paper does not pursue an absolute balance between modalities on the surface but instead aims to guide the model to more intelligently assess the intrinsic value of different modalities through the synergistic effects of gradient adjustment and feature compression. Extensive experiments demonstrate that this method achieves state-of-the art results in imbalanced feature fusion and most model robustness tasks, and validates the rationale behind the modality enhancement approach and the contribution calculation method. The contributions of this paper are as follows:

\begin{itemize}
    \item Proposed a unified paradigm for addressing modality imbalance and data noise through intelligent information compression.
    \item Proved the effectiveness of the modality enhancement approach and introduced a valid modality contribution calculation method.
\end{itemize}

\section{Related Work}
\subsection{Multi-modal Learning}

Multi-modal learning, as a complex learning paradigm, aims to integrate information from different modalities and explore the relationships between them. Current research mainly focuses on data augmentation, functional collaboration between modules, and the regulation of gradient updates for different modalities. Lin et al.\cite{lin2024adapt} explored the enhancement of multi-modal learning using Mixup data augmentation; in terms of model design, Zhou et al.\cite{zhou2025triple} improved the attention mechanism to make the model more focused on the relationships between modalities; Li et al.\cite{li2023graphmft} employed graph structures to learn complex relationships between modalities; He et al.\cite{he2025difference} adapted Variational Autoencoders \cite{kingma2013auto} to learn joint latent distributions of different modality information. Although these methods design models from various angles with the goal of enabling the model to learn as much information from different modalities as possible, they fail to consider the effectiveness of all modality information. MLA\cite{zhang2024multimodal} proposed a novel and efficient framework that alternately optimizes each modality and the shared layers independently, cleverly avoiding modality interference during joint training, and ensuring that the shared layers adapt fairly to all modalities, thereby promoting more balanced learning and more effective cross-modal knowledge fusion. Despite the presence of more information\cite{huang2022modality}, some studies suggest that due to the differences between modalities, many multi-modal learning methods still struggle to effectively enhance performance and may even experience performance degradation due to the inconsistency between modalities.

\subsection{Modality Imbalance}

Wang et al.\cite{wang2020makes} discovered that different modalities have varying convergence rates and proposed a gradient blending method that dynamically adjusts the weight of each branch based on the modality's overfitting-generalization ratio to achieve optimal fusion of modalities. Peng et al.\cite{peng2022balanced} proposed a method that dynamically adjusts the gradient update magnitude for each modality based on the contribution ratio of the modalities. For modalities with higher contributions, the gradient is reduced to slow down their optimization, whereas for those with lower contributions, the gradient is maintained or increased to accelerate optimization, thereby compensating for weaker modalities—all without introducing additional modules. Fan et al.\cite{fan2023pmr} introduced the use of prototypes to independently evaluate and rebalance the learning process of each modality, aiming to motivate the slower learning modalities and mitigate the suppressive effects of dominant modalities. Gao et al.\cite{gao2025asymmetric} took an information-theoretic approach and used mutual information to quantify both the marginal and joint contributions of each modality, guiding gradient adjustment. These methods optimize multi-modal learning through single-modality assistance or balanced learning, but they assume that all modalities have equal importance and overlook the inherent differences in capability between modalities.

\subsection{Multimodal Robustness}

When modalities are noisy, QMF \cite{zhang2023provable} dynamically adjusts the fusion weights of each modality to address the challenges posed by low-quality multi-modal data. EUA\cite{gao2024embracing}, on the other hand, directly utilizes modality uncertainty to generate augmented samples, thereby improving the model's robustness to noise. Additionally, EUA applies the Variational Information Bottleneck \cite{alemi2016deep} to compress joint representations, preventing the introduction of redundant information caused by modality alignment. Both works introduced Gaussian and Salt-Pepper noise to the data to assess the model's robustness under noisy conditions. MLA \cite{zhang2024multimodal} tackles the modality laziness problem by alternating the optimization of each modality’s encoder, which compensates for partial modality dropout. Reza et al. \cite{reza2024robust} proposed a lightweight fine-tuning approach for pretrained multi-modal networks, using adaptive layers that learn a few parameters to adjust to missing modalities.

In the case of modality missingness, MMIN\cite{zhao2021missing} employs cascaded residual autoencoders for cross-modal imputation, predicting the missing modality. IF-MMIN\cite{zuo2023exploiting} introduces modality-invariant features to explicitly mitigate the inherent discrepancies between modalities, thus improving imputation accuracy and enhancing the model’s robustness. TATE\cite{zeng2022robust} incorporates a label encoding module that signals missing modalities in the current input. These studies progressively address the modality dropout issue in multi-modal sentiment recognition.

Regarding attacks on training gradients, Yang et al. \cite{yang2024quantifying} proposed a provable lower bound for robustness, indicating that multi-modal robustness depends on the individual modality boundaries and fusion weights. To explore model robustness, they utilized the FGM \cite{goodfellow2014explaining} and $\ell_2$ PGD \cite{madry2017towards} for adversarial training attacks.

\section{Method}
This paper proposes a Contribution-Guided Asymmetric Learning (CAL) strategy. Our method introduces a dynamic, contribution-based framework that achieves asymmetric optimization through two parallel mechanisms: asymmetric gradient modulation and asymmetric information compression. These two mechanisms are unified and guided by the modality contribution metric $ W^m$. The specific architecture is shown in Figure \ref{fig:architecture}:

\begin{figure*}[pos=t] 
    \centering
    \includegraphics[width=0.8\textwidth]{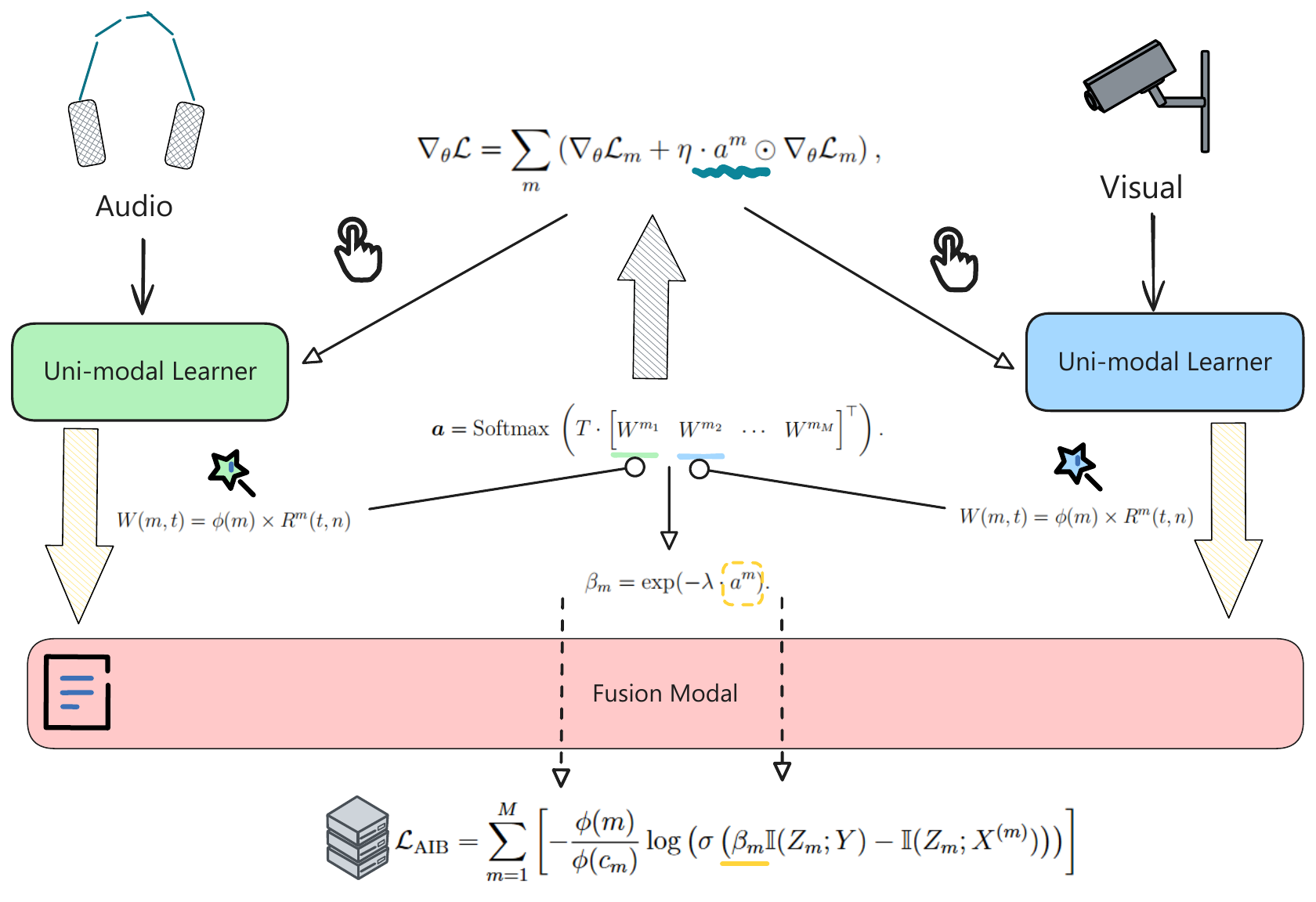} 
    \caption{The CAL architecture.}
    \label{fig:architecture}
\end{figure*}

\subsection{Modality Contribution Metric Evaluation}
The overall contribution of a modality, $ W^m $, combines both the marginal contribution and the dynamic potential of the modality within the multimodal system. The unnormalized contribution score $ W(m,t) $ is defined as the product of these two components:

\begin{equation}
\small
W(m,t) = \phi(m) \times R^m(t, n)
\label{eq:confidence_score}
\end{equation}
Where $ \phi(m) $ represents the marginal contribution of modality $ m $, and $ R^m(t) $ represents the relative performance improvement of modality $ m $ at epoch $ t $, compared to the average performance of the previous $ n $ epochs. In this paper, $ n $ is set to 5.
The larger the value of $ W(m) $, the more the model relies on modality $ m $ for decision-making, reflecting the reliability of that modality's information and its collaborative effectiveness within the multimodal system.

To simplify the computation while retaining the core idea of Shapley values \cite{shapley1953value}, which fairly evaluates the marginal contribution of a participant to a coalition, we define the marginal contribution $\phi(m)$ of modality $m$ as the average relative performance improvement brought by its inclusion in the complete modality system. Specifically, it is computed by comparing the log-likelihood difference of the prediction probabilities on the true labels between the full model (including all modalities) and a model where modality $ m $ is excluded:
\begin{equation}
\small
\begin{aligned}
\phi(m) = & \frac{\sum_{i=1}^{N} \left[ \log p(y^{(i)}|x^{\mathcal{M},(i)}) - \log p(y^{(i)}|x^{\mathcal{M} \setminus \{m\},(i)}) \right]}{N} 
\end{aligned}
\label{eq:marginal_contribution}
\end{equation}

Here, $ x^{m, (i)} $ denotes the features of modality $ m $ for the $ i $-th sample, $ y^{(i)} $ is the corresponding true label, $ p(y^{(i)}|x^{\mathcal{M}, (i)}) $ represents the prediction probability of the sample $ i $ using the fusion model with the complete set of modalities $ \mathcal{M} $, and $ \mathcal{M} \setminus \{m\} $ denotes the fusion model using all modalities except modality $ m $.

The value of $ \phi(m) $ quantifies the unique information provided by modality $ m $ and its contribution to the overall prediction performance. A larger $ \phi(m) $ indicates a higher marginal contribution of modality $ m $. This method avoids the complex subset enumeration problem in traditional Shapley value computation, requiring only the training of one full model and $ M $ ablation models, significantly reducing computational overhead.

This study proposes a potential evaluation framework based on temporal performance data, which quantifies the performance potential of modalities through three core steps: exponential smoothing, trend calculation, and residual potential estimation. This framework prevents the continuous dominance of strong modality contributions while suppressing weak modalities.

To mitigate random fluctuations in the raw performance data, we apply exponential smoothing to denoise the original sequence. The smoothed performance value $ \hat{P}^m(t) $ is defined as the weighted average of the current observed value $ P^m(t) $ and the smoothed value from the previous time step $ \hat{P}^m(t-1) $:
\begin{equation}
\small
    \hat{P}^m(t) = \gamma P^m(t) + (1 - \gamma) \hat{P}^m(t-1)
    \label{eq:smooth_performance}
\end{equation}
Where the initial condition is set as $ \hat{P}^m(0) = P^m(0) $, and $ P^m(t) $ represents the prediction score of modality $ m $ at epoch $ t $. The smoothing factor $ \gamma $ controls the smoothing intensity: the smaller the value of $ \gamma $, the smoother the sequence, but the response to recent changes becomes slower. This method improves sequence stability by adjusting the weight of historical and current data, facilitating subsequent trend analysis.

The intrinsic potential reflects the relative rate of change in modality performance, calculated by the first-order difference ratio of the smoothed data:
\begin{equation}
\small
    T^m(t,n) = \frac{ \hat{P}^m(t) - \hat{P}^m(t-1) }{ \frac{1}{n} \sum_{k=t-n}^{t-1} \hat{P}^m(k) }
    \label{eq:trend}
\end{equation}
A small constant $ \epsilon $ is introduced in the denominator, with a value of $10^{-8}$, to prevent division by zero and ensure numerical stability. The trend value $T^m(t)$ indicates performance improvement or decline based on its sign, while the absolute value represents the magnitude of the change. This design avoids distortion in the trend value due to excessively low performance baselines. The parameter $n$ defines the time window length for calculating the average performance baseline, serving as a critical parameter for balancing trend sensitivity with stability. Smaller values of $n$ respond more quickly to recent performance changes but may be more susceptible to noise interference.

The residual potential measures the future possible performance improvement of a modality. This study adopts a weighted linear model, considering both the current performance gap to the theoretical maximum and the dynamic trend reflecting the current development momentum:
\begin{equation}
\small
    R^m(t, n) = \lambda \left( P_{\text{max}} - \hat{P}^m(t) \right) + (1 - \lambda) T^m(t,n)
    \label{eq:remaining_potential}
\end{equation}
$P_{\text{max}} - \hat{P}^m(t)$ represents the static performance improvement space, with $P_{\text{max}} = 1.0$ in this paper, corresponding to the accuracy upper limit for classification tasks. The balance parameter $ \lambda $ serves as a weighting coefficient to adjust the relative importance of the static gap and dynamic trend in the final potential evaluation. This model balances the effects of static space and dynamic changes.

\subsection{Gradient Imbalance in Multimodal Learning}

In multimodal optimization, there is often a significant imbalance in gradient signals across different modalities. Some modalities dominate parameter updates in the early stages of training, while weak modalities, due to smaller gradient magnitudes or biased directions, lead to feature degradation and insufficient modality collaboration. To mitigate such optimization biases, this study proposes an adaptive gradient modulation mechanism based on modality contribution of $ W^m $, which dynamically balances the optimization strength and learning speed of different modalities.

The high-contribution modality in this paper is not necessarily the "strong" modality but rather a modality with greater potential and performance, which naturally deserves more attention. By introducing dual dependencies for contribution and asymmetric modulation, we ensure that gradient updates are matched to the actual utility of each modality, thereby improving the robustness and convergence efficiency of the fusion model.

Consider a multimodal system with $ M $ modalities. Its overall loss function can be decomposed as:
\begin{equation}
\small
\mathcal{L}_{\text{Total}} = \sum_{m=1}^{M}  \mathcal{L}_m(\theta_m, \theta_s)
\end{equation}
Where $ \theta_m $ represents the exclusive parameters of modality $ m $, and $ \theta_s $ represents the shared fusion module parameters. Based on the contribution $ W^m $, an asymmetric modulation coefficient vector $ \boldsymbol{a} = [a^{m_1}, a^{m_2}, \dots, a^{m_M}] $ is designed, which is dynamically scaled using the Softmax function with a temperature coefficient $ T $:
\begin{equation}
\small
\boldsymbol{a} = \text{Softmax}\ \left(T \cdot 
 \begin{bmatrix} W^{m_1} & W^{m_2} & \cdots & W^{m_M} \end{bmatrix}^\top 
\right)
\end{equation}
The temperature coefficient $ T $ controls the modulation sensitivity: when $ T \to 0 $, the modulation approaches a one-hot vector, reinforcing the dominant modality; when $ T \to \infty $, the modulation approaches a uniform distribution, smoothing gradient differences. Similar to the temperature parameter in contrastive learning, $ T $ enables high-contribution modalities to receive more updated driving force in the early stages of training, while low-contribution modalities gradually strengthen in the later stages, thus alleviating the learning imbalance across modalities and promoting stable convergence and robust optimization for the entire model.

During backpropagation, the gradient of the shared fusion module $g_s = \nabla_{\theta_s} \mathcal{L}_{\text{Total}} $ is modulated and then distributed to the individual modality encoders. The modulated gradient update rule is:
\begin{equation}
\small
\nabla_\theta \mathcal{L} = \sum_{m} \left( \nabla_\theta \mathcal{L}_m + a^{m} \odot \nabla_\theta \mathcal{L}_m \right)
\end{equation}
where $ \odot $ denotes element-wise multiplication. This design retains the base gradient term $ \nabla_\theta \mathcal{L}_m $to prevent feature degradation and introduces the modulation term $\eta \cdot a^{m} \odot \nabla_\theta \mathcal{L}_m$  to adjust the update magnitude of each modality. 

This design essentially forms a contribution-aware gradient update mechanism, where the modality contribution is explicitly incorporated into the gradient modulation process. It enables dynamic reallocation of optimization signals, allowing high-contribution modalities to receive more update driving force early in training, while low-contribution modalities gradually strengthen, thereby dynamically adjusting the learning rate and update magnitude across modalities and promoting the stable convergence and robust optimization of the fused feature space.

By assigning higher weight coefficients $ a^m $ to modalities with greater overall contributions, gradient descent can be effectively adjusted. For weak modalities that need to be emphasized, this coefficient can be given higher update priority, while for already important modalities, their performance can be further enhanced.

The advantage of this approach lies in its direct targeting of the optimization goal and its adaptability. The weight $ a^m $ is not fixed but is dynamically adjusted based on the contribution that each modality can provide during the training process. When a weak modality undergoes information bottleneck compression and its contribution $ W_i $ increases, reflecting a greater potential for that modality in the current optimization stage, its weight $ a^m $ is also increased, gradually transitioning towards the high-contribution modality.

\subsection{Contribution-Guided Asymmetric Information Bottleneck Compression}

In the multimodal learning framework $ (\mathcal{X} = \{ x^{(m)} \}_{m=1}^M, Y ) $, the contribution of each modality $ w_m $ to the task $ Y $ is heterogeneous. To achieve asymmetric information compression, we propose the Contribution-Guided Asymmetric Information Bottleneck (AIB) framework. This framework adjusts the information compression strength based on the modality’s contribution $ w_m $, allowing high-contribution modalities to retain more information with lower compression rates, while low-contribution modalities are compressed more strongly. We define the compression ratio as follows:
\begin{equation}
\small
\label{eq:compression_ratio}
\beta_m = \exp( -\lambda \cdot a^m)
\end{equation}
Where $ \lambda > 0 $ is a sensitivity hyperparameter used to control the strength of the impact of the weights on the compression ratio. A larger $ \lambda $ will result in more pronounced differences in compression ratios between modalities of varying importance.

In this framework, we perform information compression using a modality-specific information bottleneck approach. The features of each modality are compressed through a MLP to obtain latent representations $ z_m $.

Given two continuous random variables $ X $ and $ Y $, the mutual information based on the definition of entropy is:
\begin{equation}
\small
\label{eq:MI-diff-entropy}
I(X;Y) = H(X) + H(Y) - H(X,Y)
\end{equation}
Where $ H(\cdot) $ denotes differential entropy. This formula indicates that mutual information is the sum of the individual uncertainties of $ X $ and $ Y $ minus their joint uncertainty.

For a $ D $-dimensional Gaussian distribution $ \mathcal{N}(\boldsymbol{\mu}, \boldsymbol{\Sigma}) $, its differential entropy has a closed-form expression:
\begin{equation}
\small
\label{eq:Gaussian-entropy}
H(\mathcal{N}) = \frac{1}{2}\log\!\bigl[(2\pi e)^{D}\det(\boldsymbol{\Sigma})\bigr]
\end{equation}
Where $ \boldsymbol{\Sigma} $ is the covariance matrix and $ \det(\cdot) $ denotes the determinant. Substituting the equation \eqref{eq:Gaussian-entropy} into equation \eqref{eq:MI-diff-entropy}, we obtain the analytical expression for the mutual information between Gaussian distributed variables:
\begin{equation}
\small
I(X;Y)= \frac{1}{2}\log\frac{\det(\boldsymbol{\Sigma}_X)\,\det(\boldsymbol{\Sigma}_Y)}{\det(\boldsymbol{\Sigma}_Z)}
\label{eq:MI-gauss-final}
\end{equation}

By combining the classical Information Bottleneck principle, the optimization goal is to minimize the mutual information between the latent representation $ z $ and the input $ x $, while maximizing the mutual information between $ z $ and fused modality representation $ f_m $. The compression function in this work can be expressed as:
\begin{equation}
\small
\begin{aligned}
\mathcal{L}_{\text{AIB}} = & \sum_{m=1}^M \left[ \frac{-\log\left(\sigma\left({\beta_m {I}(Z_m; f_m) - {I}(Z_m; X^{(m)})}\right)\right) }{\text{det}_{\text{con}}} \right]
\end{aligned}
\end{equation}

Where $ \beta_m $ is the compression factor for each modality, reflecting the modality's contribution. By introducing the asymmetric compression mechanism, higher-contribution modalities are assigned lower compression strength, while lower-contribution modalities undergo stronger compression.

To ensure that the compressed features $ c_m $ maintain or enhance their original predictive utility, we introduce a marginal contribution alignment term. Using the marginal contribution $ \phi(\cdot) $, we constrain the marginal contribution $ \phi(c_m) $ of the compressed features to be close to the marginal contribution $ \phi(m) $ of the original features:
\begin{equation}
\small
\label{eq:contrib-loss}
\mathcal{L}_{\text{con}} = \left( \phi(m) - \phi(c_m) \right)^2
\end{equation}
The calculation of $ \phi(c_m) $ is the same as that of $ \phi(m) $, but with the fusion model's input replaced by the compressed features $ c_m $. This loss term prevents the compression process from destroying the unique information critical to the task.

Here, $ \phi(c_m) $ is the marginal contribution of the compressed features $ c_m $, and $ \phi(m) $ is the marginal contribution of the original features $ m $. To simplify the calculation, we define $ \text{det}_{\text{con}} = \phi(c_m) / \phi(m) $. In this way, the loss function is adjusted according to the ratio of the contributions between the compressed and original features, ensuring that the model does not lose important task-related information during compression.

To further optimize the multimodal learning process, we propose the CAL framework, which unifies the constraints of inter-modality information, fusion effectiveness, and unimodal performance within a single framework. The final optimization objective is:

\begin{equation}
\small
\mathcal{L}_{\text{Total}} = \mathcal{L}_{\text{CE}}(p^f, y) + \sum_{m \in \mathcal{M}} \mathcal{L}_{\text{CE}}(p^m, y) + \lambda_m \mathcal{L}_{\text{AIB}}
\end{equation}
Where $ \mathcal{M} = \{m_0, ..., m_i\} $ denotes the set of modalities, $ p^m $ is the output of modality $ m $, $ \lambda_m $ is a global hyperparameter for regulation, and $ \mathcal{L}_{\text{CE}} $ is the cross-entropy loss. Through the CAL framework, the model forms an adaptive, asymmetric convergence path during optimization, improving the robustness and discriminative efficiency of multimodal fusion.

\section{Experiments}

This study evaluates the proposed CAL strategy on five benchmark datasets. These datasets cover a variety of tasks including emotion recognition, event localization, action recognition, and scene recognition, providing rich multimodal data that effectively tests the multimodal learning ability across different tasks.

CREMA-D\cite{cao2014cremad} is an audiovisual dataset focused on emotion recognition. It contains six emotion categories: happiness, sadness, anger, fear, surprise, and neutrality. CREMA-D consists of a total of 7,442 video samples, with 6,698 used for training and 744 for testing. Each video sample includes facial expressions, speech, and spoken content of actors, making it effective for evaluating emotion recognition models across different modalities.

The AVE\cite{tian2018audio} dataset is specifically designed for multimodal event localization tasks. It contains 4,143 video samples covering 28 event categories. The main objective of this dataset is to evaluate the event classification ability of models in a multimodal environment, especially the accuracy and robustness of the models when combining visual and auditory information. The AVE dataset provides an important platform for evaluating event localization capabilities and is widely used in video analysis and event detection research.

Kinetics-Sounds\cite{arandjelovic2017look} is a large audiovisual dataset primarily focused on human action recognition. It contains approximately 19,000 video clips, covering 34 human action categories. Of these, 15,000 videos are used for training, 1,900 for validation, and 1,900 for testing. The Kinetics-Sounds dataset provides a high-quality multimodal action recognition resource, using both visual and audio information, and is widely applied in human behavior recognition and action detection.

NYU Depth V2\cite{silberman2012indoor} is an indoor scene recognition dataset specifically designed for multimodal learning based on RGB images and depth images. It contains 10 main scene categories, such as bedroom, kitchen, living room, etc., and is used to assess models' ability to recognize indoor scenes. The NYU Depth V2 dataset provides both RGB images and corresponding depth images, helping researchers evaluate how to combine different modalities of input to improve scene recognition accuracy in indoor environments.

MVSA\_Single\cite{niu2016sentiment} is an image-text dataset used for sentiment analysis. The MVSA\_Single dataset includes 1,555 training samples, 518 validation samples, and 519 test samples. Each sample consists of an image and corresponding text, and the task is to fuse the image and text information for sentiment classification. This dataset is primarily used for multimodal sentiment analysis and aims to study how to effectively combine image and text information to enhance sentiment classification accuracy.

\subsection{Experimental Setup and Implementation Details}

To ensure fair comparison in the experiments, baseline encoders corresponding to different tasks were selected. In the comparison experiments for imbalanced tasks (Section 4.3), mainstream imbalance strategies such as OGM\cite{peng2022OGM} and ARL\cite{Wei2025ARL} were used as baselines, with ResNet18 as the encoder backbone for the CREMA-D, AVE, and KS datasets. In the robustness attack experiments (Section 4.4), QMF and EAU\cite{gao2024embracing} were selected as baselines. For the image modality in NYU and MVSA, ResNet series models pretrained on ImageNet were used; for the text modality in MVSA, the pretrained BERT model was used. Using diverse backbone networks helps validate the generalization capability of the CAL method.

All experiments were implemented on an NVIDIA RTX 4090 GPU using PyTorch. The training configuration follows the standard settings of ARL and QMF, including a mini-batch size of 64. The image modality used the SGD optimizer with a momentum of 0.9, while the text modality used the Adam optimizer. The initial learning rate was set to $1 \times 10^{-3}$, weight decay was set to $1 \times 10^{-4}$, the smoothing coefficient $\gamma$ was set to 0.2, the window size K was set to 5, the dynamic potential parameter $\lambda$ was set to 0.5, the compression ratio parameter $\lambda$ was set to 2, and the compression parameter $\lambda_m$ was set to 0.1.

\subsection{Comparison Experiment on Imbalanced Modal Fusion}

To validate the performance of the proposed CAL method on standard imbalanced datasets, we compared it with several baseline fusion methods and imbalance learning methods, such as ARL, PMR, MLA, D\&R, on the CREMA-D, AVE, and KS datasets. Table~\ref{tab:comparison_table} shows the performance comparison of each method on the three datasets.

\begin{table}[htbp]
\centering
\small
\caption{Performance comparison of fusion and imbalance methods on CREMA-D, KS, and AVE datasets}
\label{tab:comparison_table}
\renewcommand{\arraystretch}{1.2} % 增加行高提升可读性
\setlength{\tabcolsep}{5pt} % 微调列间距
\begin{tabularx}{0.5\textwidth}{@{} >{\raggedright\arraybackslash}p{3.2cm} *{3}{>{\centering\arraybackslash}X} @{}}
\toprule
Methods & CREMA-D & KS & AVE \\
\midrule
\multicolumn{4}{c}{\textbf{Fusion Methods}} \\
\midrule
Audio-only & 57.27 & 48.67 & 62.16 \\
Visual-only & 62.17 & 52.36 & 31.40 \\
Concat & 58.83 & 64.97 & 66.15 \\
Block\cite{ben2019block} & 61.92 & 66.57 & 67.24 \\
\midrule
\multicolumn{4}{c}{\textbf{Imbalance Learning Methods}} \\
\midrule
Grad-Blending\cite{wang2020makes} & 68.81 & 67.31 & 67.40 \\
OGM-GE\cite{peng2022OGM} & 64.34 & 66.35 & 65.62 \\
AGM\cite{li2023boosting} & 67.21 & 65.61 & 64.50 \\
PMR\cite{fan2023pmr} & 65.12 & 65.01 & 63.62 \\
MMPareto\cite{wei2024mmpareto} & 70.19 & 69.13 & 68.22 \\
MLA\cite{zhang2024multimodal} & 73.21 & 69.62 & 70.92 \\
D\&R\cite{wei2024diagnosing} & 73.52 & 69.10 & 69.62 \\
ARL\cite{Wei2025ARL} & 76.61 & 74.28 & 72.89 \\
\midrule
\textbf{Ours} & \textbf{79.30} & \textbf{74.82} & \textbf{74.21} \\
\bottomrule
\end{tabularx}
\end{table}

\textbf{Experimental Results and Analysis.} 
From Table~\ref{tab:comparison_table}, it can be seen that the CAL method achieved the best performance on all three datasets: 79.30\% on CREMA-D, 74.82\% on KS, and 74.21\% on AVE, outperforming existing fusion and imbalance learning methods.

Compared to existing state-of-the-art imbalance learning methods, the CAL method also demonstrates outstanding performance. Compared to the strongest current imbalance learning method, ARL, CAL improves by 2.7\% on CREMA-D, 0.5\% on KS, and 1.3\% on AVE. While methods like ARL and MLA show certain performance in multimodal learning, they still have limitations. These methods fail to fully utilize modality contribution information to guide feature fusion, resulting in an inability to fully leverage the complementary information from different modalities when facing complex imbalanced data. In contrast, the CAL method introduces an innovative contribution-guided mechanism, which dynamically and adaptively adjusts the importance of each modality. This ensures that all modalities receive sufficient enhancement, effectively mitigating the negative impact of data imbalance.

Compared to traditional fusion methods, the CAL method shows significant performance advantages. It achieves 79.30\% on CREMA-D, a 17.4\% improvement over Block; 74.82\% on KS, an 8.3\% improvement over Block; and 74.21\% on AVE, a 7.0\% improvement over Block. This demonstrates that simple feature concatenation or block-based fusion cannot effectively address the imbalance issue between modalities. Traditional fusion methods lack a dynamic adjustment mechanism for modality contributions, which leads to strong modalities dominating the fusion process and weak modalities' information being diluted, thus limiting the upper bound of fusion performance.

\textbf{Theoretical Analysis of Method Advantages.} The advantages of the CAL method can be theoretically explained from three dimensions. The newly proposed contribution calculation method implicitly expresses the performance advantages of modalities through marginal contribution. It does not solely rely on performance but also incorporates the calculation of contribution potential to prevent excessive focus on high-performance modalities. Through the asymmetric information bottleneck compression, the CAL method can explicitly compress redundant information from modalities while preserving the task-specific features unique to each modality, resulting in more compact and effective fused feature representations, and effectively encouraging the improvement of weak-contribution modalities. Through the dynamic gradient modulation strategy, the CAL method adaptively adjusts the learning intensity of each modality based on its contribution, ensuring that the key information from strong-contribution modalities is fully utilized, while preventing weak modalities from being completely suppressed and retaining their task-related useful information, achieving more balanced and effective multimodal learning.

Experimental results show that the CAL method's outstanding performance across multiple datasets validates the effectiveness of its design. Compared to unimodal methods, the proposed multimodal information fusion results in significant performance improvements, proving the advantage of multimodal learning in uncovering complementary information. Compared to existing imbalance learning methods, the proposed method achieves better modality alignment and feature compression, thus demonstrating stronger robustness and higher accuracy in imbalanced data scenarios.

\subsection{Robustness Comparison Experiments}

To verify the robustness and adaptability of the model in noisy environments, we conducted comprehensive noise robustness experiments on the MVSA-Single and NYU Depth v2 datasets. Two noise scenarios were designed for the experiments. The first is a generalization test, where noise is applied only to the test set to evaluate the model's generalization ability to unknown noise. The second is a learning ability test, where noise is applied to both the training and test sets to assess the model’s learning ability in a noisy environment. Two types of noise, salt-and-pepper noise and Gaussian noise, were used with different noise intensities $\epsilon=5$ and $\epsilon=10$, to comprehensively evaluate the model’s performance under varying noise conditions.

\textbf{Generalization and Learning Ability Experimental Results.} Table~\ref{tab:combined_noisy_results_image_style} shows the comparison of the generalization ability of each method in a noisy environment. The results indicate that the CAL method demonstrates significant robustness advantages in noisy environments. On the MVSA-Single dataset, under light noise conditions, the CAL method achieves an accuracy of 76.02\% under salt-and-pepper noise and 76.83\% under Gaussian noise, which is an improvement of 1.2\% and 2.9\%, respectively, over the EAU method. Under heavy noise conditions, the CAL method achieves an accuracy of 69.75\% under salt-and-pepper noise and 63.74\% under Gaussian noise, improving by 8.7\% and 1.7\%, respectively, compared to the EAU method, indicating stronger noise generalization ability. On the NYU Depth v2 dataset, under light noise, the CAL method achieves 61.12\% and 63.01\% accuracy under salt-and-pepper noise and Gaussian noise, respectively, improving by 1.3\% under salt-and-pepper noise compared to the EAU method, while performance is roughly the same under Gaussian noise. Under heavy noise, the CAL method achieves 48.63\% and 59.06\% accuracy under salt-and-pepper noise and Gaussian noise, respectively, improving by 1.8\% and 0.2\% over the EAU method. In contrast, although methods like EAU perform well on clean data, their performance significantly degrades in noisy environments, reflecting their relative lack of noise adaptation capability.

Table~\ref{tab:performance_comparison} shows the comparison of the learning ability of each method when noise is applied to both the training and test sets. The results show that the CAL method exhibits certain advantages in learning ability in noisy environments. On the MVSA-Single dataset, under light noise, the CAL method achieves an accuracy of 77.61\% under salt-and-pepper noise and 78.41\% under Gaussian noise, improving by 4.2\% and 6.0\%, respectively, over the EAU method. Under heavy noise, the CAL method achieves an accuracy of 76.62\% under salt-and-pepper noise and 77.99\% under Gaussian noise, improving by 5.5\% and 6.6\%, respectively, compared to the EAU method. On the NYU Depth v2 dataset, under light noise, the CAL method achieves 67.43\% and 66.52\% accuracy under salt-and-pepper noise and Gaussian noise, respectively, improving by 1.1\% and 1.5\% compared to the EAU method. Under heavy noise, the CAL method achieves 69.11\% and 64.83\% accuracy under salt-and-pepper noise and Gaussian noise, respectively, improving by 2.0\% and 0.5\% over the EAU method. Even under higher noise intensities, the CAL method is able to maintain good performance, outperforming other comparison methods. These results show that the CAL method not only excels in noise generalization but also exhibits strong adaptability in noisy learning environments.

\begin{table}[H]
\centering
\small
\caption{The comparison results of model generalization robustness in the absence of noise environment. Noise perturbations were applied only to the test set to evaluate the generalization ability of each method to unknown noise.}
\label{tab:combined_noisy_results_image_style}
\begin{tabular}{l r r r r r}
\toprule
\multicolumn{6}{c}{\textbf{Noisy MVSA-Single}} \\
\midrule
\multirow{2}{*}{Method} & \multicolumn{1}{c}{Clean} & \multicolumn{2}{c}{Salt-Pepper Noise} & \multicolumn{2}{c}{Gaussian Noise} \\
\cmidrule(lr){2-2} \cmidrule(lr){3-4} \cmidrule(lr){5-6}
 & \multicolumn{1}{c}{$\epsilon = 0$} & \multicolumn{1}{c}{$\epsilon = 5$} & \multicolumn{1}{c}{$\epsilon = 10$} & \multicolumn{1}{c}{$\epsilon = 5$} & \multicolumn{1}{c}{$\epsilon = 10$} \\
\midrule
Concat & 65.59 & 58.69 & 51.16 & 50.70 & 46.12 \\
Bert\cite{devlin2019bert} & 75.61 & 69.50 & 47.41 & 69.50 & 47.41 \\
MMBT\cite{kiela2019supervised} & 78.50 & 74.07 & 51.26 & 71.99 & 55.35 \\
TMC\cite{han2022trusted} & 74.87 & 68.02 & 56.62 & 66.72 & 60.36 \\
QMF\cite{zhang2023provable} & 78.07 & 73.90 & 60.41 & 73.85 & 61.28 \\
EAU\cite{gao2024embracing} & \textbf{79.15} & 74.81 & 61.04 & 73.89 & 62.04 \\
\textbf{Ours} & 78.32 & \textbf{76.02} & \textbf{69.75} & \textbf{76.83} & \textbf{63.74} \\
\midrule\midrule 
\multicolumn{6}{c}{\textbf{Noisy NYU Depth v2}} \\
\midrule
\multirow{2}{*}{Method} & \multicolumn{1}{c}{Clean} & \multicolumn{2}{c}{Salt-Pepper Noise} & \multicolumn{2}{c}{Gaussian Noise} \\
\cmidrule(lr){2-2} \cmidrule(lr){3-4} \cmidrule(lr){5-6}
 & \multicolumn{1}{c}{$\epsilon = 0$} & \multicolumn{1}{c}{$\epsilon = 5$} & \multicolumn{1}{c}{$\epsilon = 10$} & \multicolumn{1}{c}{$\epsilon = 5$} & \multicolumn{1}{c}{$\epsilon = 10$} \\
\midrule
Concat & 70.44 & 57.98 & 44.51 & 59.97 & 53.20 \\
Align\cite{jia2021scaling} & 70.31 & 57.54 & 43.01 & 59.47 & 51.74 \\
MMTM\cite{joze2020mmtm} & 71.04 & 59.45 & 44.59 & 60.37 & 52.28 \\
TMC\cite{han2022trusted} & 71.01 & 59.34 & 44.65 & 61.04 & 53.36 \\
QMF\cite{zhang2023provable} & 70.06 & 58.50 & 45.69 & 61.62 & 55.60 \\
EAU\cite{gao2024embracing} & \textbf{72.05} & 59.83 & 46.85 & \textbf{63.33} & 58.85 \\
Ours & 71.16 & \textbf{61.12} & \textbf{48.63} & 63.01 & \textbf{59.06} \\
\bottomrule
\end{tabular}
\end{table}

\textbf{Analysis of Results.} The asymmetric information compression mechanism improves the model's generalization ability. AIBLOSS dynamically adjusts the information compression strength based on modal contribution, applying stronger compression to weaker modalities, thereby promoting their gradual enhancement and increasing their attention. As a result, even when strong modalities are attacked during the testing phase, the model can maintain high robustness. Additionally, the compression mechanism helps better concentrate the distribution of feature information, effectively filtering out redundant information and reducing interference.

\begin{table}[H]
\centering
\caption{Results of Model Learning Ability in Noisy Training Environment. The experiment applies noise perturbations to both the training and test sets, evaluating the adaptability of each method in noisy environments.}
\label{tab:performance_comparison}
\begin{adjustbox}{max width=\linewidth}
\begin{tabular}{l c c c c c}
\toprule
\multirow{2}{*}{Method} & & \multicolumn{2}{c}{Salt-and-Pepper Noise} & \multicolumn{2}{c}{Gaussian Noise} \\
\cmidrule(lr){3-4} \cmidrule(lr){5-6}
 & $\varepsilon = 0.0$ & $\varepsilon = 5.0$ & $\varepsilon = 10.0$ & $\varepsilon = 5.0$ & $\varepsilon = 10.0$ \\
\midrule
\multicolumn{6}{c}{\textbf{NYU Depth V2}} \\ % 修改1: 将横跨列数改为6，并使用左对齐加粗文本
\midrule
Concat & 70.44 & 60.08 & 47.24 & 60.02 & 55.27 \\
QMF\cite{zhang2023provable} & 70.06 & 65.07 & 65.58 & 64.62 & 62.54 \\
EAU\cite{gao2024embracing} &   \textbf{72.05} & 66.35 & 67.13 & 65.01 & 64.30 \\
\textbf{Ours} & 71.12 & \textbf{67.43} & \textbf{69.11} & \textbf{66.52} & \textbf{64.83} \\
\midrule
\multicolumn{6}{c}{\textbf{MVSA-Single}} \\ % 修改2: 同上
\midrule
 Concat  & 64.08 & 54.12 & 48.77 & 53.26 & 50.45 \\
 QMF\cite{zhang2023provable} & 78.07 & 76.64 & 74.21 & 76.52 & 75.87 \\
 EAU\cite{gao2024embracing}   & \textbf{79.15} & 73.45 & 71.12 & 72.38 & 71.40 \\
 \textbf{Ours} & 77.92 & \textbf{77.61} & \textbf{76.62} & \textbf{78.41} & \textbf{77.99} \\
\bottomrule
\end{tabular}
\end{adjustbox}
\end{table}

The learning advantage in noisy environments primarily stems from its dynamic adaptation mechanism. Through the contribution-guided mechanism, the model can dynamically identify modalities affected by noise during training. When a modality is contaminated by noise, its marginal contribution $\phi(m)$ and performance potential $R^m(t, n)$ both decrease, leading to a reduction in the contribution weight of that modality, which naturally suppresses the impact of the noise. This dynamic weight adjustment mechanism allows the model to adaptively handle different levels of noise without the need for explicit noise detection or denoising preprocessing.

Additionally, through the dynamic gradient modulation strategy, the learning intensity of each modality is adaptively adjusted based on its contribution. In noisy environments, modalities with higher robustness and contribution typically receive higher gradient weights, thus being updated preferentially, while the learning intensity of low-contribution modalities, which contain more noise, is reduced, thereby preventing noise from interfering with the overall learning process. The asymmetric information compression method also effectively filters out redundant information, retaining useful discriminative features. This dynamic adaptation mechanism allows the CAL method to achieve more stable learning in noisy environments, and it is easier to learn robust feature representations when the training data also contains noise.

\subsection{Ablation Experiment}

\subsubsection{Impact of AIB Loss on Modal Imbalance Adjustment and Robustness}

Table \ref{tab:ablations_AVE_CREMA} shows the ablation experiment results of AIB Loss on the NYUD2 and CREMA-D datasets, focusing on its role in modal imbalance adjustment and robustness. This experiment removes the gradient impact to specifically analyze the performance of AIB Loss under different configurations.

In the configuration using minimal modality information "mi", feature compression is performed by minimizing the mutual information between the compressed features and the original features. However, due to the lack of adjustment of the compression coefficient and guidance in the "mx" direction, the compressed features failed to effectively focus on removing redundant information, and instead deviated from the effective information during the blind compression process, resulting in a decrease in accuracy.

In contrast, after introducing the maximum modality information "mx" configuration, the model maximizes the mutual information between the compressed features and the fused features, allowing the target features to focus effectively on key information. Unlike the blind compression in the "mi" configuration, the "mx" configuration, although still lacking the adjustment of the compression coefficient for each modality, still compresses towards the correct direction of the fused features with less redundancy, thus achieving a significant performance improvement. Particularly, under salt-and-pepper noise intensities $\varepsilon = 5.0$ and $\varepsilon = 10.0$, accuracy reached 58.78\% and 46.69\%, respectively, which is notably better than the configuration without maximum modality information, further validating the effectiveness of the "mx" configuration in optimizing multimodal feature compression.

The configuration combining both minimal and maximal modality information "mx\&mi" allows the target compressed features to escape from the redundant parts of the original features while compressing towards the key information direction of the fused features. This effectively achieves reasonable compression and enhancement of modality information, further improving the model's robustness under different noise conditions. On the NYUD2 dataset, this configuration achieved 59.90\% and 45.54\% accuracy at $\varepsilon = 5.0$ and $\varepsilon = 10.0$, respectively, which is an improvement over the accuracy of the single "mx" configuration.

In the modality contribution-weighted $\beta$ configuration, under the correct guidance of "mx\&mi" compression, the model fully leverages the advantages of high-contribution modalities over low-contribution modalities. The compression strength of the latter modality is increased to remove redundant information, while high-contribution modalities retain more key information, ensuring that the feature representations of all modalities converge towards task-relevant core information. Specifically, by dynamically adjusting modality contributions, AIB Loss can effectively guide the information compression process, reducing redundant information and enhancing feature alignment across modalities. Notably, at $\varepsilon = 5.0$ and $\varepsilon = 10.0$, the model's accuracy reached 61.10\% and 48.62\%, respectively, showing higher stability and robustness compared to other configurations, proving the unique advantage of AIB Loss in enhancing multimodal model performance and handling heterogeneous information. Moreover, the $\frac{1}{\beta}$ configuration, although able to adjust modality contributions to some extent, did not perform better than the standard $\beta$ configuration. This confirms the critical role of compression strength in optimizing multimodal models.

In summary, the performance of AIB Loss under different configurations verifies its advantages in modal imbalance adjustment and robustness enhancement. The "mx\&mi" configuration effectively focuses each modality on key information while removing redundant information, improving the model's performance in both clean data and noisy environments. The optimal $\beta$ configuration, by evaluating the contribution of each modality, dynamically adjusts the compression strength of different modal features. This configuration allows for more precise removal of redundant information and guides each modality towards a task-relevant semantic space, enabling the model to show higher stability and robustness when faced with multimodal information imbalance.
\begin{table}[H]
\centering
\small
\caption{Ablation Experiment Results on the NYUD2 and CREMA-D Datasets under Different Conditions.}
\label{tab:ablations_AVE_CREMA}
\begin{adjustbox}{max width=0.5\textwidth}
% 减少列间距，使表格更紧凑
\setlength{\tabcolsep}{6pt} % 默认约为6pt，适当减小以压缩整体宽度
\begin{tabularx}{0.5\textwidth}{c@{\hskip 1pt} c c c *{3}{>{\centering\arraybackslash}X}} 
\toprule
\multicolumn{1}{c}{\multirow{2}{*}{$\beta$}} & \multicolumn{1}{c}{\multirow{2}{*}{mi}} & \multicolumn{1}{c}{\multirow{2}{*}{mx}} & \multicolumn{1}{c}{\multirow{2}{*}{mx\&mi}} & \multicolumn{2}{c}{NYUD2} & \multicolumn{1}{c}{CREMA-D} \\
\cmidrule(lr){5-6} \cmidrule(lr){7-7}
 &  &  &  & $\varepsilon = 5.0$ & $\varepsilon = 10.0$ & $\varepsilon = 0$ \\
\midrule
 & $\checkmark$ & &  & 52.54 & 42.62 & 75.92 \\
 &  & $\checkmark$ &  & 58.78 & 46.69 & 77.34 \\
 &  &  & $\checkmark$ & 59.90 & 45.54 & 77.83 \\
$1/\beta$ &  &  & $\checkmark$ & 58.35 & 47.28 & 77.81 \\
$\beta$&  &  & $\checkmark$ & 61.10 & 48.62 & 79.30 \\
\bottomrule
\end{tabularx}
\end{adjustbox}
\end{table}

\subsubsection{Analysis of the Rationality of Contribution Degree Calculation Method}

To verify the rationality of the proposed modal contribution calculation formula $ W_i = \phi(m) \times R^m(t, n) $, we conducted a comprehensive ablation experiment, systematically comparing five different contribution calculation methods: $\phi \times R$ (ours); additive combination $\phi + R$; marginal contribution $\phi$ only; performance potential $R$ only; and the method based on single-modal accuracy $P \times R$. The experiment aims to validate the advantages of $W_i = \phi(m) \times R^m(t, n)$ over other combinations from both theoretical and empirical perspectives.

We evaluated the five contribution calculation methods on the AVE dataset, using the same training configuration. All methods share the same network architecture and training process, with the only difference being the calculation of the modal contribution to ensure a fair comparison.

\begin{table}[H]
\centering
\small
\caption{Ablation Experiment Results: Accuracy Comparison of Different Contribution Calculation Methods}
\label{tab:ablations_contrib_methods}
\begin{adjustbox}{max width=0.5\textwidth}
\begin{tabularx}{0.5\textwidth}{l *{3}{>{\centering\arraybackslash}X}}
\toprule
\textbf{Method} & \textbf{FusionACC} & \textbf{AudioACC} & \textbf{VisualACC} \\
\midrule
$\phi$ only & 0.7135 & 0.6328 & 0.3047 \\
$R$ only & 0.7188 & 0.6250 & 0.2812 \\
$\phi + R$ & 0.7396 & 0.6589 & 0.3438 \\
$P\times R$ & 0.7352 & 0.6531 & 0.3376 \\
\textbf{$\phi \times R$} & \textbf{0.7421} & \textbf{0.6610} & \textbf{0.3457} \\
\bottomrule
\end{tabularx}
\end{adjustbox}
\end{table}
From Table~\ref{tab:ablations_contrib_methods}, we can see that our $\phi \times R$ method achieves the best fusion performance, outperforming all other comparison methods. It improves fusion accuracy by 2.86\%, 2.33\%, 0.25\%, and 0.69\% compared to $\phi$ only, $R$ only, $\phi + R$, and $P \times R$, respectively. This advantage can be explained from the perspectives of Shapley value theory and dynamic balancing mechanisms. The multiplicative relationship in the modal contribution $ W_i = \phi(m) \times R^m(t, n) $ establishes a necessary condition constraint mechanism, where the modal contribution will only significantly improve when both the marginal contribution $\phi(m)$ and performance potential $R^m(t, n)$ are at high levels. This design aligns with the fundamental requirements of multimodal learning, where an effective modality should not merely offer a unique marginal contribution but also exhibit a clear potential for sustained performance gains. 
From the perspective of joint probability, the overall quality of a modality can be seen as the joint event of marginal contribution and performance potential, and the probability properties of joint events determine that multiplication is the most appropriate combination. Compared to the $P \times R$ method based on single-modal accuracy, our $\phi \times R$ method, by calculating the marginal contribution $\phi(m)$ using Shapley value theory, can more accurately evaluate the unique value of the modality in the multimodal system.

The methods $\phi$ only or $R$ only do not achieve ideal performance levels. The $\phi$ only method considers only the marginal contribution of the modality, completely ignoring the dynamic performance changes of the modality during the training process, which may result in selecting a modality with high marginal contribution but stagnant performance. Even if a modality has a high marginal contribution, if its performance improvement is slow or stagnant during the training process, it may indicate that the modality has reached a performance bottleneck, and further reliance on this modality may not lead to additional performance improvements. On the other hand, the $R$ only method allocates weights based solely on the relative performance improvement of the modality during the training process, ignoring its actual marginal contribution in the multimodal system. This may lead to selecting modalities with rapid performance gains but limited marginal contribution, where the modality shows a good performance improvement trend during training, but the unique information it provides has limited marginal contribution to the overall prediction performance and may be redundant with other modalities.

Although the $\phi + R$ method performs better than the individual metric methods, it still significantly underperforms the $\phi \times R$ method. The additive relationship $W_i = \phi(m) + R^m(t, n)$ lacks necessary condition constraints. It allows the weight to remain high even when one of the metrics is low and the other is high, which does not meet the practical needs of multimodal learning, where modalities must both have significant marginal contributions and show sustained performance improvements. In contrast, the multiplicative relationship$W_i = \phi(m) \times R^m(t, n)$inherently enforces a simultaneous satisfaction requirement due to its product form: the product attains a high value only when both factors are large, whereas if one factor is small, the product diminishes significantly even if the other remains high, thereby allowing for more precise discrimination of the most valuable modalities.

\begin{figure}[pos=t] 
    \centering    
    \begin{minipage}{0.48\columnwidth}
        \centering
        \includegraphics[width=\linewidth]{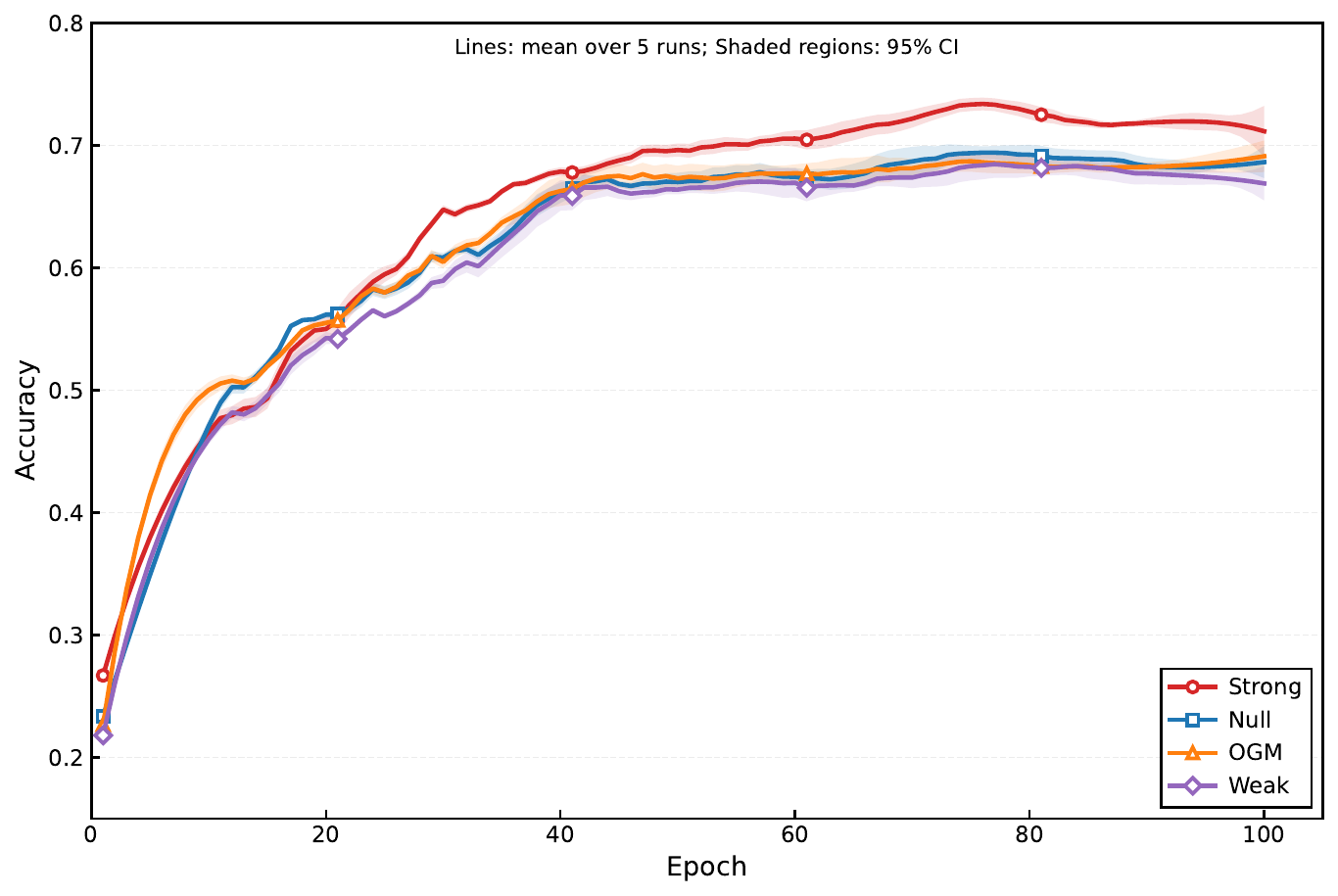}
    \end{minipage}%
    \hfill
    \begin{minipage}{0.48\columnwidth}
        \centering
        \includegraphics[width=\linewidth]{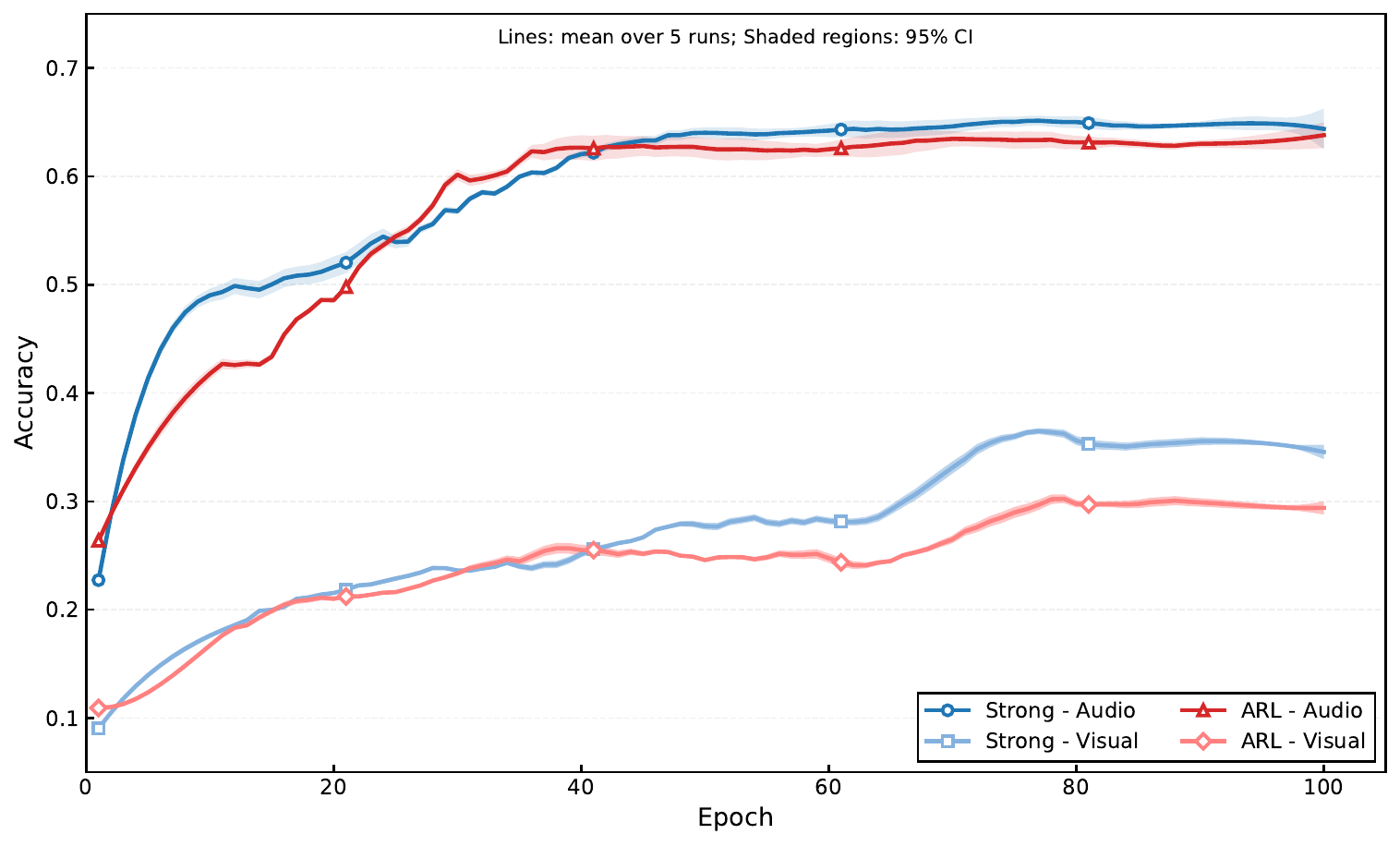}
    \end{minipage}
    \caption{Left: Performance of the four gradient modulation strategies (Strong, None, OGM, Weak) on the ACC-epoch curve. Right: Comparison between the Strong strategy and ARL strategy for the audio and visual modalities.}
    \label{fig:acc_epoch_combined}
\end{figure}

\subsubsection{Gradient Modulation Strategy Validation}

To validate the effectiveness of the proposed gradient modulation strategy, we conducted comprehensive comparison experiments on the AVE dataset, evaluating four different gradient modulation strategies using the same training configuration. The gradient modulation mechanisms are as follows: the Strong mode uses additive gradient increases, where the weights are calculated based on the modality's contribution degree, favoring modalities with higher contributions; the Weak mode uses inverted weights to strengthen the low-contribution modalities; the None mode does not apply any gradient modulation, with fixed weights of $[0.5, 0.5]$; the OGM mode uses gradient subtraction to suppress the strong modality.

Table~\ref{tab:gradient_ablation} shows the comparison of fusion accuracy across the four methods. From the results, the Strong strategy achieved the highest fusion accuracy of 74.21\%, outperforming the other three methods. The OGM mode reached 72.45\%, slightly higher than the None mode at 72.18\%, while the Weak mode performed the worst, with an accuracy of only 70.95\%. The Strong mode improved by 1.76\% compared to OGM, 2.03\% compared to None, and 3.26\% compared to Weak. This indicates that strengthening the gradient modulation for high-contribution modalities can effectively improve multimodal fusion performance.

\begin{table}[H]
\centering
\caption{Gradient modulation strategy comparison results on the AVE dataset.}
\label{tab:gradient_ablation}
\begin{tabular}{lcc}
\toprule
Method & Fusion Accuracy & Relative Difference from Strong \\
\midrule
\textbf{Strong} & \textbf{74.21} & \textbf{best} \\
OGM & 72.45 & $-$1.76 \\
None & 72.18 & $-$2.03 \\
Weak & 70.95 & $-$3.26 \\
\bottomrule
\end{tabular}
\end{table}

Figure~\ref{fig:acc_epoch_combined} on the left shows the performance of the Strong, None, OGM, and Weak strategies on the ACC-epoch curve, with the shaded area representing the 95\% confidence interval. From the visualization, it is clear that the Strong mode maintained stable and efficient performance improvement throughout the training process. By dynamically adjusting the gradient weights, it achieved the best multimodal alignment and fusion quality, a result that is also verified by the t-SNE visualization.

The OGM mode is slightly higher than the None mode but significantly lower than the Strong mode. From Figure~\ref{fig:acc_epoch_combined}, it can be seen that the performance curve of the OGM strategy lies below that of the Strong strategy, which verifies that although the gradient subtraction strategy to suppress strong modalities is better than the inverted strategy, it is still not as effective as the strategy of strengthening high-contribution modalities through gradient addition. The None strategy lacks a dynamic gradient adjustment mechanism and cannot fully leverage the complementary information between modalities. While it outperforms the Weak strategy, it is still lower than both the Strong and OGM modes. The Weak strategy performed the worst, and from the left side of Figure~\ref{fig:acc_epoch_combined}, it can be seen that the inverted weight strategy led to the performance curve remaining at the lowest level, affecting learning stability and verifying the detrimental effect of the inverted strategy.

The right side of Figure~\ref{fig:acc_epoch_combined} shows a comparison of the dynamic gradient weight changes between the Strong strategy and the ARL strategy for the audio and visual modalities. From the right plot, it can be observed that the Strong strategy achieved better modal weight distribution during training through dynamic gradient modulation. Specifically, this method can adaptively adjust the weights based on the modality's contribution degree, giving higher gradient weights to high-contribution modalities while maintaining moderate weights for weaker modalities, thereby achieving asymmetric but coordinated multimodal alignment.

Compared to gradient suppression modulation, gradient enhancement modulation can more effectively balance the update magnitudes of the gradients for each modality during the optimization process, maintaining a smooth and continuous optimization trend in training. As shown in the left panel of Figure~\ref{fig:acc_epoch_combined}, the Strong mode, using additive gradient modulation, has a smooth and continuously rising performance curve, ultimately reaching the optimal point. In contrast, the subtractive OGM strategy may weaken the gradient of key information during the gradient counteraction process, which can lead to training fluctuations and performance bottlenecks. This suggests that CAL enhances the gradient signal of high-contribution modalities through additive modulation, improving model convergence stability while promoting robust overall performance improvement.

Experimental results show that the Strong mode achieves optimal feature alignment and redundancy compression through reasonable gradient weight distribution, significantly outperforming other methods. This verifies our hypothesis that enhancing high-contribution modalities creates a better asymmetric alignment mechanism for multimodal learning. These visual and quantitative evidences collectively demonstrate the important impact of the gradient modulation strategy on multimodal learning performance.

\subsubsection{Feature Alignment Visualization}

To validate the effectiveness of AIBLOSS in compressing redundant features and promoting multimodal alignment, we performed a t-SNE visualization analysis on the CREMAD dataset, comparing three configurations: an untrained model, a baseline model without AIBLOSS, and a model with AIBLOSS enabled. To ensure meaningful comparison, we adopted a consistent t-SNE transformation strategy by concatenating the 512-dimensional raw features from all models and performing a joint projection to ensure consistency in the coordinate system.

\begin{figure*}[pos=t]
\centering
\includegraphics[width=\textwidth]{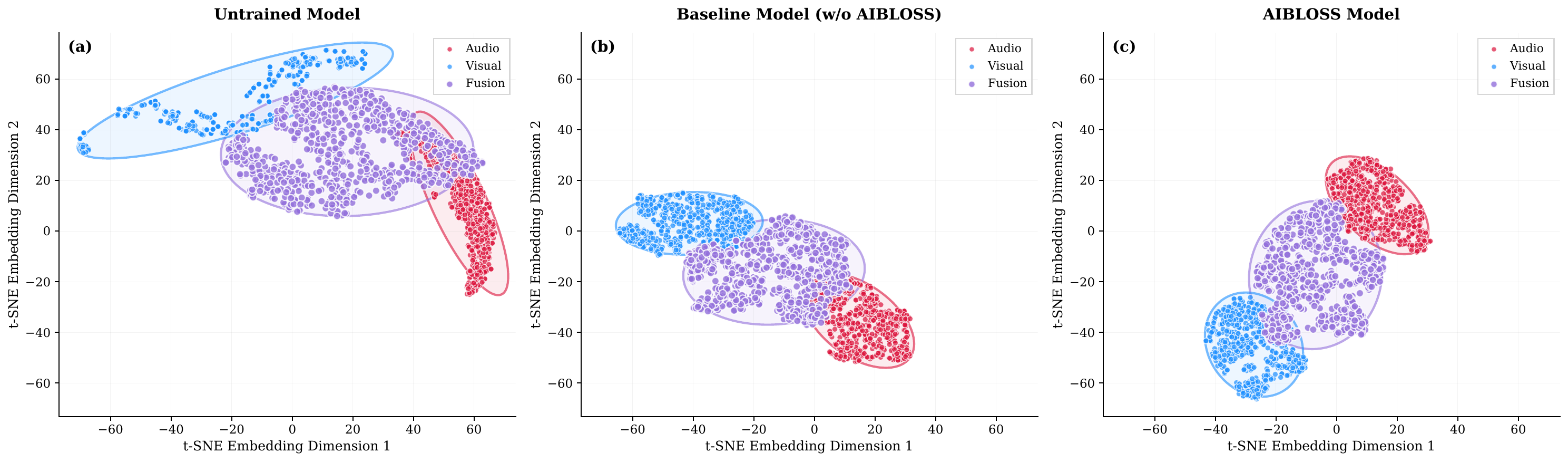}
\caption{t-SNE feature distribution visualization for three model configurations on the CREMAD dataset. (a) Untrained model. (b) Baseline model without AIB LOSS. (c) AIB LOSS model. The ellipses represent the 95\% confidence regions for each feature type. All visualizations share a unified coordinate system for direct comparison.}
\label{fig:tsne_comparison}
\end{figure*}

Figure~\ref{fig:tsne_comparison} shows the evolution of feature distributions under the three model configurations. In the untrained model shown in Figure~\ref{fig:tsne_comparison}(a), audio and visual features are clearly separated in the embedding space, with minimal overlap, reflecting that no effective association has been established between modalities in the random initialization state. The fused features are scattered randomly without any clear localization.

Figure~\ref{fig:tsne_comparison}(b) presents the baseline model, showing some improvement, where the audio and visual clusters are closer together and the overlap area is larger. However, there is still a noticeable separation between the two modality clusters, and the fused features are not precisely located at the intersection center. The feature clusters for each modality remain relatively dispersed, indicating that redundant modality-specific information has not been effectively compressed.

In contrast, our model, shown in Figure~\ref{fig:tsne_comparison}(c), exhibits significant alignment, with both audio and visual features converging towards the fused feature, forming a tightly overlapping triplet of feature distributions. The fused features are precisely located at the intersection center of the two modality clusters, indicating that AIB LOSS has successfully established the fused features as alignment anchors. The confidence ellipses are tighter compared to both the untrained model and the baseline model, suggesting that redundant information has been effectively compressed while preserving complementary discriminative features and task-specific features unique to each modality.

The visualization results validate the effectiveness of AIB LOSS in compressing redundant features and promoting multimodal alignment. The bidirectional convergence towards the fused feature indicates that the asymmetric information compression strategy creates a natural alignment mechanism, using the fused features as anchors to guide the two modalities to discard redundant modality-specific information while moving towards a shared discriminative feature space. This geometric shift confirms our core hypothesis: explicit feature compression through asymmetric information compression helps modalities focus on task-critical information.

\subsubsection{Visualization of Contribution Calculation Method}

To validate the effectiveness of the proposed modality contribution $W_i$, we analyzed the dynamic relationship between modality contribution and accuracy on the CREMA-D dataset. We tracked the changes in modality contributions and their corresponding accuracy variations during the training process, analyzing the correlation between contribution and performance metrics.

\begin{figure}[pos=t]
    \centering
    \includegraphics[width=0.45\linewidth]{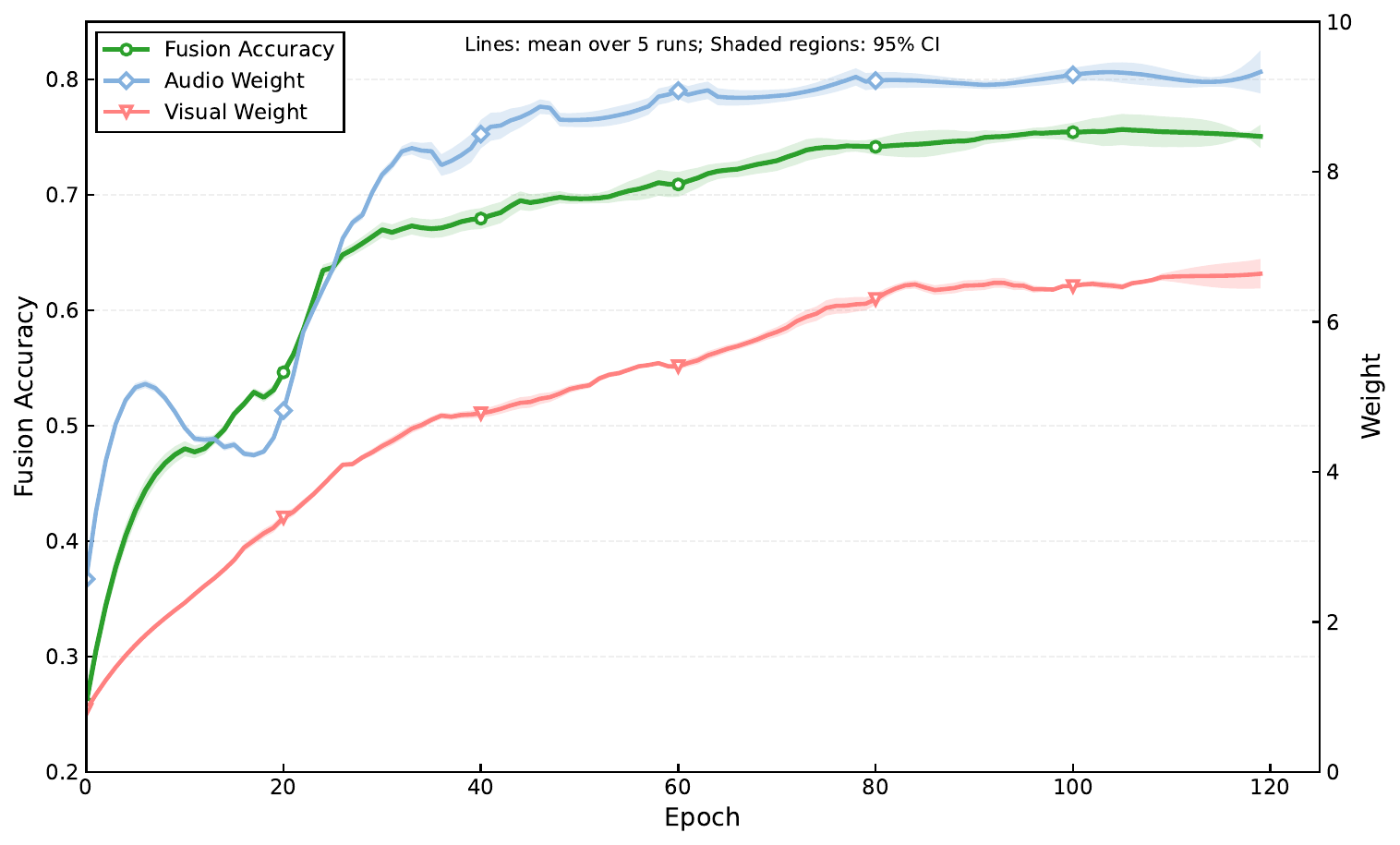}
    \hspace{0.04\linewidth}
    \includegraphics[width=0.45\linewidth]{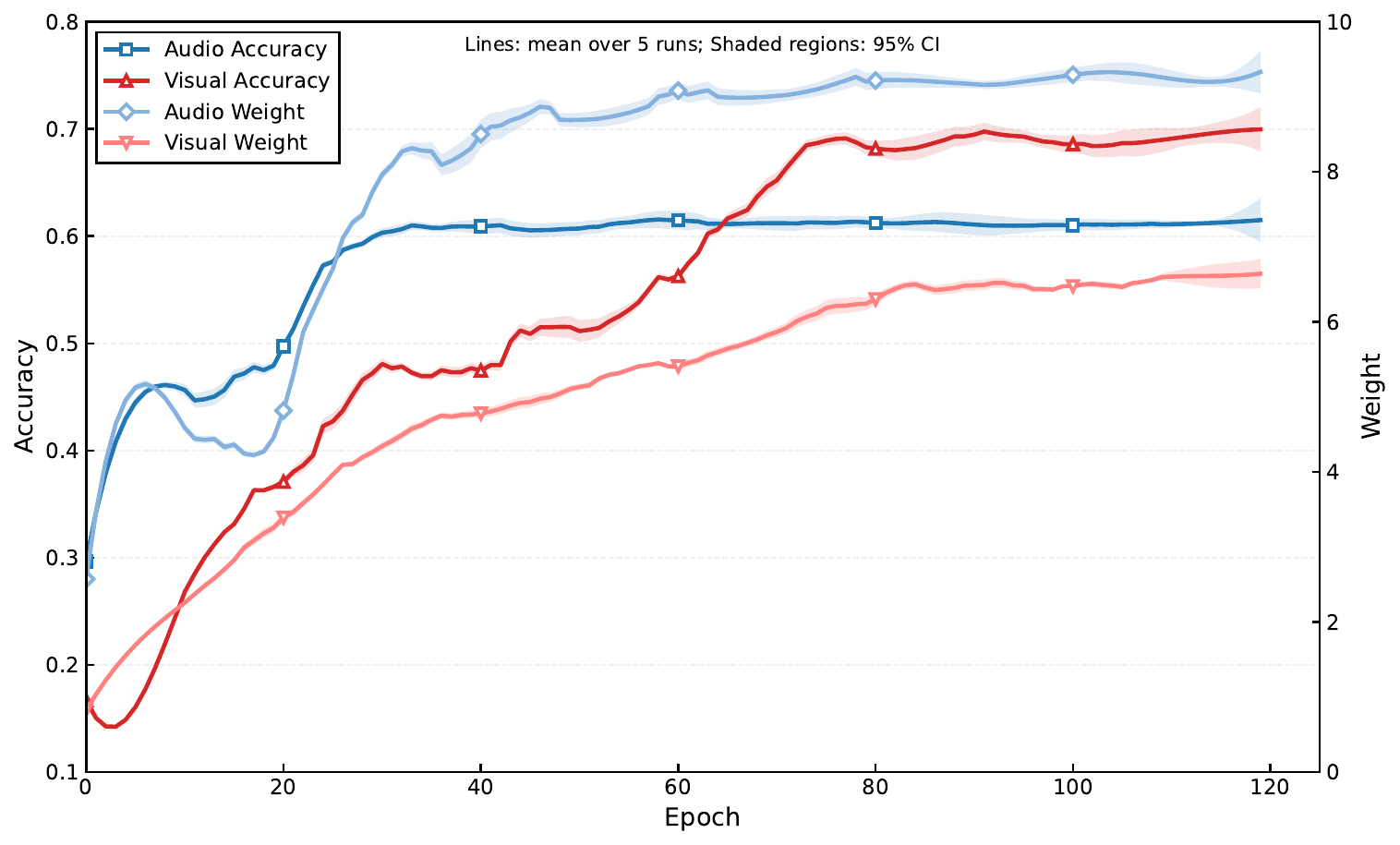}
    \caption{Left: Relationship between single-modality accuracy and its contribution. Right: Relationship between fusion accuracy and modality contribution. The shaded area indicates the 95\% confidence interval.}
    \label{fig:contribution_validation}
\end{figure}

Figure~\ref{fig:contribution_validation}(a) shows the relationship between the accuracy of individual modalities and their corresponding contributions, while Figure~\ref{fig:contribution_validation}(b) shows the relationship between fusion accuracy and modality contribution. From the experimental results, the following key phenomena can be clearly observed:

From Figure~\ref{fig:contribution_validation}(a), it can be seen that the accuracy of the audio modality is significantly higher than that of the visual modality in the early stages of training. Correspondingly, the contribution of the audio modality is relatively higher in the initial stages of training. This phenomenon thus justifies the definition of the marginal contribution $\phi(m)$, as it confirms that a modality's assigned contribution should correlate with its marginal value. The marginal contribution of a single modality reflects its prior importance to the task, and using it as a factor in contribution calculation is reasonable. Notably, as training progresses, the two contribution curves intersect multiple times, reflecting the dynamic change in modality identity. When the accuracy of the visual modality increases rapidly, its contribution temporarily exceeds that of the audio modality, validating the effectiveness of the dynamic adjustment mechanism of performance potential $R^m(t, n)$.

Coordination of Performance Potential and Fusion Performance. From Figure~\ref{fig:contribution_validation}(b), it can be seen that as training progresses and fusion accuracy steadily improves, the distribution of modality contributions also stabilizes and becomes more reasonable. The contribution of the audio and visual modalities maintains a relatively stable ratio throughout the training process, and this contribution distribution is positively correlated with the improvement in fusion accuracy. This indicates the effectiveness of performance potential $R^m(t, n)$, which prevents the uncontrolled enhancement of strong modalities and suppresses weak modalities, leading to better fusion results.

The Necessity of Joint Consideration. Considering only the marginal contribution $\phi(m)$ may lead to the suppression of weak modalities, while considering only the performance potential $R^m(t, n)$, may fail to fully leverage the prior advantages of the modality. Our formula $W_i = \phi(m) \times R^m(t, n)$ combines the two orthogonal factors organically through a multiplicative form, ensuring that the distribution of contributions respects the prior importance of the modality while also considering its actual contribution to multimodal fusion. As shown in Figure~\ref{fig:contribution_validation}, this joint consideration approach results in a more reasonable distribution of contributions, highly consistent with the changing trend of performance metrics, while allowing for dynamic changes in modality identity during training, better adapting to the modality characteristics at different stages of training.

\section{Conclusion}
\label{sec:conclusion}

The proposed CAL framework addresses the challenges of optimization imbalance and noise interference in multi-modal learning. In contrast to conventional approaches that primarily suppress dominant modalities, CAL introduces a novel contribution metric derived from both marginal performance and dynamic potential. This metric adaptively reinforces high-contribution modalities and guides asymmetric gradient updates to maximize the utility of the most informative sources .Concurrently, the framework employs asymmetric compression on modality features, which effectively reduces noise while simultaneously promoting the training of weaker modalities. This dual mechanism enhances overall model robustness against noise and alleviates modality imbalance.

Empirical evaluations on five benchmark datasets demonstrate that the proposed framework achieves state-of-the-art results in both imbalanced fusion and noise-affected scenarios, outperforming strong baselines such as ARL, QMF and EAU. Ablation studies confirm the superiority of the dynamic reinforcement strategy over strong-modality suppression and validate the efficacy of the contribution calculation method based on $\phi(m) \times R^m(t, n)$. Designed as a flexible and modular framework, it can be readily transferred to other multi-modal tasks.

Despite these advancements, the current contribution assessment mechanism relies on manually tuned parameters, which may introduce subjective bias. Consequently, developing more intelligent, adaptive methods for evaluating modality contribution and adjusting gradients remains a crucial direction for further refining the framework.

\bibliographystyle{elsarticle-num}
\bibliography{ref}

% Generated by IEEEtran.bst, version: 1.14 (2015/08/26)
\begin{thebibliography}{10}
\providecommand{\url}[1]{#1}
\csname url@samestyle\endcsname
\providecommand{\newblock}{\relax}
\providecommand{\bibinfo}[2]{#2}
\providecommand{\BIBentrySTDinterwordspacing}{\spaceskip=0pt\relax}
\providecommand{\BIBentryALTinterwordstretchfactor}{4}
\providecommand{\BIBentryALTinterwordspacing}{\spaceskip=\fontdimen2\font plus
\BIBentryALTinterwordstretchfactor\fontdimen3\font minus \fontdimen4\font\relax}
\providecommand{\BIBforeignlanguage}[2]{{%
\expandafter\ifx\csname l@#1\endcsname\relax
\typeout{** WARNING: IEEEtran.bst: No hyphenation pattern has been}%
\typeout{** loaded for the language `#1'. Using the pattern for}%
\typeout{** the default language instead.}%
\else
\language=\csname l@#1\endcsname
\fi
#2}}
\providecommand{\BIBdecl}{\relax}
\BIBdecl

\bibitem{wang2020makes}
W.~Wang, D.~Tran, and M.~Feiszli, ``What makes training multi-modal classification networks hard?'' in \emph{Proceedings of the IEEE/CVF conference on computer vision and pattern recognition}, 2020, pp. 12\,695--12\,705.

\bibitem{huang2022modality}
Y.~Huang, J.~Lin, C.~Zhou, H.~Yang, and L.~Huang, ``Modality competition: What makes joint training of multi-modal network fail in deep learning?(provably),'' in \emph{International conference on machine learning}.\hskip 1em plus 0.5em minus 0.4em\relax PMLR, 2022, pp. 9226--9259.

\bibitem{cao2014cremad}
H.~Cao, D.~G. Cooper, M.~K. Keutmann, R.~C. Gur, A.~Nenkova, and R.~Verma, ``Crema-d: Crowd-sourced emotional multimodal actors dataset,'' \emph{IEEE transactions on affective computing}, vol.~5, no.~4, pp. 377--390, 2014.

\bibitem{peng2022OGM}
X.~Peng, Y.~Wei, A.~Deng, D.~Wang, and D.~Hu, ``Balanced multimodal learning via on-the-fly gradient modulation,'' in \emph{Proceedings of the IEEE/CVF conference on computer vision and pattern recognition}, 2022, pp. 8238--8247.

\bibitem{Wei2025ARL}
S.~Wei, C.~Luo, and Y.~Luo, ``Improving multimodal learning via imbalanced learning,'' in \emph{Proceedings of the IEEE/CVF International Conference on Computer Vision (ICCV)}, October 2025, pp. 2250--2259.

\bibitem{gao2024embracing}
Z.~Gao, X.~Jiang, X.~Xu, F.~Shen, Y.~Li, and H.~T. Shen, ``Embracing unimodal aleatoric uncertainty for robust multimodal fusion,'' in \emph{Proceedings of the IEEE/CVF conference on computer vision and pattern recognition}, 2024, pp. 26\,876--26\,885.

\bibitem{yang2024quantifying}
Z.~Yang, Y.~Wei, C.~Liang, and D.~Hu, ``Quantifying and enhancing multi-modal robustness with modality preference,'' \emph{arXiv preprint arXiv:2402.06244}, 2024.

\bibitem{zhao2021missing}
J.~Zhao, R.~Li, and Q.~Jin, ``Missing modality imagination network for emotion recognition with uncertain missing modalities,'' in \emph{Proceedings of the 59th Annual Meeting of the Association for Computational Linguistics and the 11th International Joint Conference on Natural Language Processing (Volume 1: Long Papers)}, 2021, pp. 2608--2618.

\bibitem{zeng2022robust}
J.~Zeng, J.~Zhou, and T.~Liu, ``Robust multimodal sentiment analysis via tag encoding of uncertain missing modalities,'' \emph{IEEE Transactions on Multimedia}, vol.~25, pp. 6301--6314, 2022.

\bibitem{li2023graphmft}
J.~Li, X.~Wang, G.~Lv, and Z.~Zeng, ``Graphmft: A graph network based multimodal fusion technique for emotion recognition in conversation,'' \emph{Neurocomputing}, vol. 550, p. 126427, 2023.

\bibitem{ben2019block}
H.~Ben-Younes, R.~Cadene, N.~Thome, and M.~Cord, ``Block: Bilinear superdiagonal fusion for visual question answering and visual relationship detection,'' in \emph{Proceedings of the AAAI conference on artificial intelligence}, vol.~33, no.~01, 2019, pp. 8102--8109.

\bibitem{lin2024adapt}
R.~Lin and H.~Hu, ``Adapt and explore: Multimodal mixup for representation learning,'' \emph{Information Fusion}, vol. 105, p. 102216, 2024.

\bibitem{zhou2025triple}
Y.~Zhou, X.~Liang, H.~Chen, Y.~Zhao, X.~Chen, and L.~Yu, ``Triple disentangled representation learning for multimodal affective analysis,'' \emph{Information Fusion}, vol. 114, p. 102663, 2025.

\bibitem{he2025difference}
C.~He, S.~Song, Z.~Jia, and H.~Zhao, ``Difference bonds consistency and complementarity to enhance multimodal representation learning,'' in \emph{ICASSP 2025-2025 IEEE International Conference on Acoustics, Speech and Signal Processing (ICASSP)}.\hskip 1em plus 0.5em minus 0.4em\relax IEEE, 2025, pp. 1--5.

\bibitem{kingma2013auto}
D.~P. Kingma and M.~Welling, ``Auto-encoding variational bayes,'' \emph{arXiv preprint arXiv:1312.6114}, 2013.

\bibitem{peng2022balanced}
X.~Peng, Y.~Wei, A.~Deng, D.~Wang, and D.~Hu, ``Balanced multimodal learning via on-the-fly gradient modulation,'' in \emph{Proceedings of the IEEE/CVF conference on computer vision and pattern recognition}, 2022, pp. 8238--8247.

\bibitem{gao2025asymmetric}
X.~Gao, B.~Cao, P.~Zhu, N.~Wang, and Q.~Hu, ``Asymmetric reinforcing against multi-modal representation bias,'' in \emph{Proceedings of the AAAI Conference on Artificial Intelligence}, vol.~39, no.~16, 2025, pp. 16\,754--16\,762.

\bibitem{zuo2023exploiting}
H.~Zuo, R.~Liu, J.~Zhao, G.~Gao, and H.~Li, ``Exploiting modality-invariant feature for robust multimodal emotion recognition with missing modalities,'' in \emph{ICASSP 2023-2023 IEEE International Conference on Acoustics, Speech and Signal Processing (ICASSP)}.\hskip 1em plus 0.5em minus 0.4em\relax IEEE, 2023, pp. 1--5.

\bibitem{reza2024robust}
M.~K. Reza, A.~Prater-Bennette, and M.~S. Asif, ``Robust multimodal learning with missing modalities via parameter-efficient adaptation,'' \emph{IEEE Transactions on Pattern Analysis and Machine Intelligence}, 2024.

\bibitem{goodfellow2014explaining}
I.~J. Goodfellow, J.~Shlens, and C.~Szegedy, ``Explaining and harnessing adversarial examples,'' \emph{arXiv preprint arXiv:1412.6572}, 2014.

\bibitem{madry2017towards}
A.~Madry, A.~Makelov, L.~Schmidt, D.~Tsipras, and A.~Vladu, ``Towards deep learning models resistant to adversarial attacks,'' \emph{arXiv preprint arXiv:1706.06083}, 2017.

\bibitem{zhang2023provable}
Q.~Zhang, H.~Wu, C.~Zhang, Q.~Hu, H.~Fu, J.~T. Zhou, and X.~Peng, ``Provable dynamic fusion for low-quality multimodal data,'' in \emph{International conference on machine learning}.\hskip 1em plus 0.5em minus 0.4em\relax PMLR, 2023, pp. 41\,753--41\,769.

\bibitem{alemi2016deep}
A.~A. Alemi, I.~Fischer, J.~V. Dillon, and K.~Murphy, ``Deep variational information bottleneck,'' \emph{arXiv preprint arXiv:1612.00410}, 2016.

\bibitem{zhang2024multimodal}
X.~Zhang, J.~Yoon, M.~Bansal, and H.~Yao, ``Multimodal representation learning by alternating unimodal adaptation,'' in \emph{Proceedings of the IEEE/CVF conference on computer vision and pattern recognition}, 2024, pp. 27\,456--27\,466.

\bibitem{fan2023pmr}
Y.~Fan, W.~Xu, H.~Wang, J.~Wang, and S.~Guo, ``Pmr: Prototypical modal rebalance for multimodal learning,'' in \emph{Proceedings of the IEEE/CVF Conference on Computer Vision and Pattern Recognition}, 2023, pp. 20\,029--20\,038.

\bibitem{shapley1953value}
L.~S. Shapley \emph{et~al.}, ``A value for n-person games,'' 1953.

\end{thebibliography}

% To print the credit authorship contribution details
\printcredits

%% Loading bibliography style file
%\bibliographystyle{model1-num-names}
% \bibliographystyle{cas-model2-names}

\end{document}